\documentclass[aip]{revtex4-1}
\usepackage{graphicx}
\usepackage{dcolumn}
\usepackage{bm}%
\usepackage{color}

\preprint{APS/123-QED}

\newcommand{\spineln}{$\rm Co_3O_4$}
\newcommand{\spinel}{$\rm Co_3O_4$ }

\newcommand{\inter}{$\rm  Co_3O_4/ZnO$ }
\newcommand{\coan}{$\rm Co^{2+}$}
\newcommand{\coa}{$\rm Co^{2+}$ }
\newcommand{\cobn}{$\rm  Co^{3+}$}
\newcommand{\cob}{$\rm  Co^{3+}$ }

\begin{document}
\title{Interface magnetism and electronic structure: ZnO(0001)/Co\textsubscript{3}O\textsubscript{4}(111)} 

\author{I.M. Kupchak}
\author{N.F. Serpak}
\affiliation{
V.Lashkarev Institute of Semiconductor Physics, NAS Ukraine, 
45, Pr. Nauky, Kyiv, 03680, Ukraine
}
\email{kupchak@isp.kiev.ua}
\author{A. Shkrebtii}
\affiliation{
University of Ontario, Institute of Technology, 
2000 Simcoe Street North, Oshawa, Ontario L1H 7K4, Canada
}
\author{R. Hayn}
\affiliation{
Aix-Marseille Universit\'e, CNRS, IM2NP-UMR 7334, 13397 Marseille Cedex 20, France 
}



\date{\today}

\begin{abstract}

We have studied the structural, electronic and magnetic properties of spinel \spinel (111) surfaces and their interfaces with ZnO (0001) 
using density functional theory (DFT) within the Generalized Gradient Approximation with on-site Coulomb repulsion term (GGA+U). 
Two possible forms of spinel surface, containing \coa or \cob ions and terminated with either cobalt or oxygen ions were 
considered, as well as their interface with zinc oxide. Our calculations demonstrate that \cob ions attain non-zero magnetic moments at the 
surface and interface, in contrast to the bulk, where they are not magnetic, leading to the ferromagnetic ordering. Since heavily 
Co-doped ZnO samples can contain \spinel secondary phase, such a magnetic ordering at the interface might explain the origin of the magnetism in such 
diluted magnetic semiconductors (DMS).

\end{abstract}

\maketitle

\tableofcontents

\section{ Introduction}

Magnetic semiconductors (MS) and diluted magnetic semiconductors (DMS) exhibit both ferromagnetic and semiconducting properties. Therefore, they  are 
promising materials for spintronics, which utilizes for information processing not only the electron charge but also its spin. Historically, 
the first DMS with a high Curie temperature up to about 200~K was GaAs doped with Mn ions.\cite{1,2} In that compound, the ferromagnetism 
is promoted by hole carriers, which align along the local Mn magnetic moments and called carrier-induced ferromagnetism or Zener $p-d$ exchange. It is crucial 
for this mechanism that Mn at the Ga site becomes $\rm Mn^{2+}$ instead of $\rm Ga^{3+}$, thus providing at the same time a local spin and a hole 
charge carrier. Extension of the mechanism, proposed in a very influential paper \cite{3} of Dietl and co-workers, allows a prediction that the above room-temperature ferromagnetism in ZnO:Co and 
GaN:Mn is due to the same carrier-induced mechanism. This would be responsible for the ferromagnetism 
with a sufficiently high number of hole charge carriers. First 
experiments after that prediction \cite{4} seemed to confirm the mechanism proposed and has also been  supported by \textit{ab-initio} calculations.\cite{5} However, 
it soon turned out that the Co impurity is in fact isovalent to the Zn ion \cite{6} and provides no charge carriers at all,  while the  situation 
in GaN:Mn is similar.\cite{7}

We are going to concentrate here on ZnO:Co, where the experimental reports  demonstrate that the above room-temperature 
ferromagnetism in ZnO:Co persist. Even though its origin is still not clarified, there are clear indications in more recent experiments that 
the magnetism in the ZnO:Co system is attributed to the formation of the \spinel phase in ZnO.\cite{2006Nipan,2007Wang,2009Li,2015Colak,2007Dietl} Therefore, we will focus here on the 
role of the \spinel phase, although several attempts to explain the mechanism of the
ferromagnetism in realistic ZnO:Co systems exist including, for instance, spinodal decomposition  \cite{8} or  Lieb-Mattis ferrimagnetism,\cite{9}  to 
cite just two ideas. 

The typical doping level of Co in ZnO can be relatively high (in the range between 10\% and 30\%). This leads to the secondary phases of \spinel 
and $\rm ZnCo_2O_4$ segregation during the sample growth, which can be detected, for instance, by Raman spectroscopy.\cite{10,11} Although, 
in general, the appearance of such secondary phases is detrimental for the DMS materials, this effect can also be advantageous. However, a lack of understanding of the 
secondary phases and their interfaces remains currently the main obstacle toward 
the practical applications of \spinel surfaces and their interfaces. By carrying out the first principle simulations of the \inter interface we  
offer not only  the realistic explanations in the big puzzle of the nature of ferromagnetism in DMS, but also show the promise for the
new applications.

Cobalt oxide \spineln, also known as tricobalt tetraoxide or cobal spinel, is a \textit p-type semiconductor with the reported optical energy 
band gap $E_g$ between 1.1 and 1.65~eV (see [\onlinecite{12}] and ref. therein). It is widely used in lithium-ion batteries as a cathode 
material,\cite{13} gas sensing, nanomaterials and nano-junctions, environmental and numerous other applications.\cite{14,15,16,17} \spinel 
crystallizes in the cubic normal spinel structure, which contains cobalt ions in two different oxidation states, \coa and \cobn, located at 
the interstitial tetrahedral (A) and octahedral (B) sites, respectively (see., $e.g.$,  Ref.~[\onlinecite{12}]). The bulk magnetic properties of the cobalt oxide are well understood. In the presence of tetrahedral 
crystal field, the five-fold degenerate atomic {\textit d} orbitals of \coa ions are split into two groups, $e_g$ and $t_{2g}$, leading to three 
unpaired {\textit d} electrons on $t_{2g}$ orbital. Similarly, in a case of \cob ion, the crystal field is octahedral, and the splitting leads to 
six paired electrons in the $t_{2g}$ orbital, while $e_g$ orbital is empty. As a result, the \coa ions carry a permanent magnetic moment, 
whereas \cob ions are not magnetic. Considering the A-site sublattice only, each \coa ion is surrounded by four neighbors with oppositely 
directed spin, thus forming an antiferromagnetic (AFM)  state. In general, such nearest A-A exchange interaction is expected to be weak, since in the
typical spinel structures with magnetic cations A-B coupling between the ions in tetrahedral and octahedral sites is dominant.\cite{18} However, 
in the \spinel spinel this A-A coupling is unusually strong due to the indirect exchange through the intermediate \cob ions in the octahedral 
B-site, providing \coa ions by a magnetic moment of about $3.02~\mu_B$. As a result of such strong coupling, \spinel is antiferromagnetic below 
the N\'eel temperature $T_{N}\sim40~K$ and paramagnetic at higher temperatures.\cite{18}

When such a complex structure is terminated by a surface or forms an interface, one can expect new interesting magnetic peculiarities, absent in 
the bulk of the crystal. Indeed, formation of surface or interface between different materials involves several important factors such as 
surface polarity, charge transfer, stresses, defects, {\textit etc.}, altering the long-range magnetic ordering, and the magnetic 
response as a result.\cite{19,20} There are many publications on the electronic and magnetic properties of different spinels and their surfaces, 
such as $\rm Fe_3O_4$ spinel (see, {\textit e.g.}, [\onlinecite{20}]), which has the crystal structure similar to that of \spineln. However, the 
cobalt spinel surfaces are still not that well understood, while even more complex behaviour should be expected when the interface with other 
materials is formed. It has been shown that during the epitaxial growth of \spineln, two surfaces with the lowest surface energy, namely (111) and 
(110), are typically formed.\cite{21} More detailed experimental and theoretical study have been performed in [\onlinecite{22}], where the effect 
of different \spinel crystal planes orientation has been investigated. This aimed in reducing charge-discharge over-potential toward an application in high energy density  $\rm Li-O_2$ 
batteries and it was established, that (111) surface is the most efficient. Experimentally, cobalt spinel (110) surface has been thoroughly 
investigated by Petitto and Langel  \cite{23} using low energy electron diffraction (LEED), Auger electron spectroscopy (AES), and X-ray photoelectron 
spectroscopy (XPS). The \spinel (111) surface  has been studied by X-ray diffraction (XRD) and atomic force microscopy (AFM) methods,\cite{24} LEED 
and scanning tunneling microscopy (STM).\cite{25,26,27,28} Bulk \spinel have also been studied using Raman spectroscopy.\cite{29} In general, \spinel 
attracts the interest because of its high catalytic activity, especially for CO oxidation,\cite{51} therefore most of the research have been performed 
toward such an application. Concerning the theory,  a number of publications has been dedicated to {\textit ab-initio}  
study of electronic and 
magnetic properties of the bulk and surfaces of \spineln.\cite{30,31,32,33,34,35,36,Kupchak2017} The main problem, discussed in the above cited theoretical 
works, was usually a nature of superexchange in bulk spinel and the stability of its surfaces under different conditions, such as different atom 
types (\coan, \cob ions, or O) at the top layer termination.

Another field of cobalt spinel application is related to the interface between {\textit p}-type \spinel and {\textit n}-type ZnO, which 
forms {\textit p-n} heterojunction. In particular, {\textit p}-\spineln/{\textit n}-ZnO composites can provide higher sensitivities and faster 
responses toward gas sensoring application.\cite{37,38,39,40,16} Such composites are typically obtained using a mixture of ZnO and \spinel powders 
and following annealing, that forms inhomogeneous interface between both semiconductors. However, the presence of this interface also plays a 
significant role in the magnetic properties of such composites. Indeed, there is an evidence of the magnetism appearance in ZnO/\spinel powder 
mixture at room temperature even without thermal treatment.\cite{41,42} Authors explain this phenomena by surface reduction of the \spinel nanoparticles, 
in which the antiferromagnetic \spinel nanoparticle is surrounded by a CoO-like shell. Other authors,\cite{43,44} studying ZnO/\spinel powder mixture 
by X-ray absorption spectroscopy (XAS) and optical spectroscopy, explained such phenomena by reduction \cobn$\,\to\,$\coa at the \spinel nanoparticle 
surface. This explanation has been proved by Vibrating Sample Magnetometer (VSM) analysis of composite ZnO, synthesized on the surface of core \spinel 
in [\onlinecite{45}]. Recently a diode consisting of {\textit p}-type \spinel nanoplate / {\textit n}-type ZnO nanorods heteroepitaxal junction has been 
fabricated, showing reasonable electrical performance,\cite{46} but no attention has been paid to its magnetic properties. Despite of extensive investigation 
of the cobalt oxides, mentioned above, still there is no clear picture of the role of the cobalt oxide surfaces and interfaces on the magnetic properties.

Considering the lack of microscopic understanding of the surface and interface magnetism, the present study is aimed to establish the nature of ferromagnetism at 
the \inter interface toward an application in the new device types for spintronics. We have investigated from first principles modifications of the atomic structure
 at various types of the \inter boundaries, related changes in the electronic band structure and their contribution to the appearance of 
the interface magnetic properties. The paper is organized as following. We present in Section 2 the numerical formalism, which is used throughout the paper. 
Section 3 discusses the microscopic atomic structure of the \spineln(111) surfaces and  \inter interfaces. The results of the calculated magnetic and 
electronic properties and their modifications due to the surfaces or interfaces, are discussed in Section 4. The conclusion is presented in Section 5.

\section{Numerical Method}
We investigated the atomic and electronic structure of the \inter interface within the density functional theory (DFT) and generalized gradient 
approximation (GGA), as implemented in the Quantum-Espresso software package.\cite{47} We have used ultrasoft Perdew-Burke-Ernzerhof (PBE) pseudopotentials, 
\cite{48} which include 12 valence electrons for zinc, 6 valence electrons for oxygen, and 9 valence electrons for cobalt. An integration of the 
Brillouin zone has been performed using $4\times4$ $\Gamma$-centered grid of special points in {\textit k}-space, generated by Monkhorst-Pack scheme  
\cite{49} and Methfessel-Paxton smearing \cite{50} with a parameter of 0.005~Ry. Several tests were performed with denser grids up to $10\times10$, 
but no significant changes have been observed compared to the case of $4\times4$ grid. To ensure a sufficient convergence of the results we applied 
40~Ry cutoff for smooth part of the wave function and 400~Ry for the augmented charge density. 
We approximated the exchange-correlation functional with both the local spin resolved generalized gradient approximation (SGGA) and the so-called 
SGGA+U approximation, in which the effect of electron correlations in the  3{\textit d} shell is taken into account by considering the on-site 
Coulomb interactions within the Hubbard method.\cite{32,Dorado2010}
We have chosen the value of Hubbard U parameters to be 3.5~eV and 5.0~eV for Co and Zn atoms, 
respectively. 

Although the Hubbard parameters chosen are commonly accepted in the literature, they still are a subject of discussion. \cite{32} 
Therefore, DFT+U calculations of \spinel should be carried out with care: the systems under consideration might have several solutions and there is no guarantee 
whether the lowest energy solution corresponds to the global minimum. 
For such a reason, we have checked that our conclusions do not depend in a sensitive way on these 
Coulomb parameters. As discussed in [\onlinecite{Dorado2010}], the DFT+U instability can be further exacerbated in the presence of the {\textit f}-orbitals and the absence of the gap between 
the filled and empty states. However, considered here surfaces and interfaces are semiconducting 
and the {\textit f}-states are not present. To make sure that the Coulomb parameters choice does not affect our results, we followed the established approach from [\onlinecite{34}]. In particular, 
(i) we applied a Methfessel-Paxton smearing technique of the Brillouin-zone integration that, as proved, ensures the convergence to the global minimum both for metals 
and systems with nonzero energy gap. (ii) We have also considered several different values of the Hubbard parameter and found that the calculations consistently converges to the same energy. 

To optimize the atomic geometry of \spinel surfaces and \inter interfaces we have performed 
the structural relaxations within the SGGA method, while the final calculations of the magnetic structure and the densities of states has been carried out using the SGGA+U 
method. The systems were relaxed through all the internal coordinates until the Hellmann-Feynman forces became less than $10^{-4}$~a.u., while 
keeping the shape and the volume of the supercell fixed.

\section{Surface and interface structural details}
To investigate the origin of the surface/interface magnetism we model two types of \spineln(111) surfaces and two ZnO(0001)/\spineln(111) interfaces. While the above considered 
surfaces and interfaces are well suited to simulate numerically, in addition to (111) planes differently oriented interfaces were observed experimentally.\cite{21,52,37,Liu2013,39} 
However, as we discuss below, the main magnetic features, predicted for (111) system considered, should also be common for differently oriented interfaces in the experimentally observed materials. 

The bulk terminated atomic structure of \spineln(111) spinel surface in [111] direction, perpendicular to the surface, can be described by a sequence 
of atomic layers, containing \coa ions or both of \coa and \cob ions, separated by layer of oxygen: $\rm O-Co^{2+}-O-Co^{2+}Co^{3+}$. The primitive 
unit cell, containing such a sequence, has a hexagonal symmetry along the surface or the interface. The upper  layer, which forms the interface with 
ZnO, contains either three \cob ions (B-terminated layer) or a combination of two \coa and one \cob ions (A-terminated layer, such convention is used since the closest to 
the interface cobalt oxide layer is of A-type). The interface between the upper layer of \spinel and ZnO is then being formed by introducing a single 
layer of four oxygen atoms, which  match the Co--O bonds of the spinel. These four oxygen atoms can also be viewed as those belonging to ZnO in a sequence 
$\rm Zn-O-Zn-O$ of the primitive unit cell: the topology of this spinel oxygen layer has the same symmetry as (0001) plane of hexagonal ZnO. Hence, 
to form the epitaxial interface with \spinel and to saturate these oxygen bonds, four primitive unit cells of hexagonal ZnO are required. In such a way, 
oxygen atoms play a role of a ``bridge'' between cubic spinel \spinel and wurtzite ZnO. 

\begin{figure}
\includegraphics{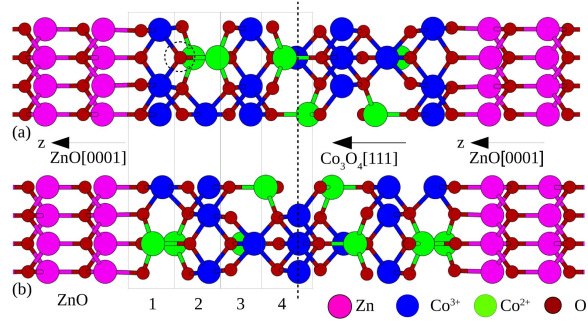} 
\caption{(Color online) Side view of the unit cells of \inter interfaces: (a) ``Octahedral'' B-terminated interface, (b) ``Tetrahedral'' A-terminated 
interface. Numbers denote the atomic layers of both interfaces.}
\label{fig:struct}
\end{figure}

We have paid special attention when choosing the lateral unit cell size of the interface for the systems under investigation since \spineln(111) and ZnO(0001) demonstrate 
considerable lattice mismatch. To simulate the ZnO(0001)/\spineln(111)  interface three possibilities exist: (i) choosing the spinel bulk constant to determine the interface unit cell size, 
(ii) using ZnO bulk parameters to define the  interface unit cell, and (iii) optimizing the lattice parameter for the interface to find the unit cell size that minimizes the total energy of the interface. 
Following the experimental finding, we did not optimize the lattice parameter for the interface, since such optimization should lead to both ZnO and \spinel material stressed. 
Indeed, for our case, while such mismatch should make the epitaxial growth of the flat \spineln/ZnO interface challenging, the experimental microscopic images demonstrate 
the smooth interface between \spinel inclusions and ZnO host material \cite{39, 52,37} without noticeable modification of the interlayer distances and dislocation appearance.
Since \spinel is supposed to be the source of the magnetism, first we have chosen spinel bulk constant as the main structural parameter, resulted in compressed ZnO part of the system, 
and then relaxed the atomic positions in the interface vicinity. Considering the experimental value of bulk spinel lattice constant $a_{spinel}=8.084$~\AA, a primitive unit cell of its (111) surface has a lattice
constant $c_{spinel}=5.72$~\AA. Since the corresponding parameter of ZnO has a value $c_{ZnO}=3.25$~{\AA}, in order to fit four primitive unit cells 
of ZnO onto single 2D unit cell of spinel, the bulk constant of ZnO should be compressed in the basal plane by about 12\%. Therefore, the lattice 
constant of this strained ZnO at the interface region is 2.86~\AA. As it has been suggested in [\onlinecite{46}] for ZnO nanorods on \spinel nanoplates, such a large stress is relieved 
by forming dislocations along basal plane at the interface. In the case of the \spinel inclusions, however, they have low-sized diameters, which allow to easily accommodate the strain through 
the lateral relaxation, thus making heteroepitaxal growth possible even in the case of high lattice mismatch.\cite{51} Additionally, in the present calculations such a 
stress effect is partially taken into account by the system relaxation within the unit cell. On the other hand, study of possible extended dislocations, originated due to the mismatch, requires simulation of 
significantly larger unit cells and was out of the scope of our research. 

Second, we have chosen to test the ZnO unit cell size for the interface, which resulted in ``stretched'' \spineln side. However, when we carried out the relaxation of the atomic positions in the interface vicinity, 
the spinel-like structure of \spinel has not been preserved. Again, considering the experimental finding of the bulk-like \spinel spinel inclusions on XRD spectra \cite{Phan2012}, existence of such stretched \spinel systems 
does not look credible. To confirm this conclusion we again compared our theoretical results with the experimental Raman spectra, as discussed below. 

Therefore, to study magnetic and electronic structures of an interface, we created two symmetric slabs, containing seven atomic layers of \spineln, and ZnO layers, adjacent on 
both sides, as shown in Fig.~\ref{fig:struct}.  The first slab (Fig.~\ref{fig:struct}a) is composed of a spinel top layer containing \cob 
ions at the B-sites only (``octahedral'' interface), while the second slab (Fig.~\ref{fig:struct}b) contains at the interface both \coa (A-site) and one \cob (B-site) 
ions (``tetrahedral'' interface). In such a way, each slab contains two interface regions of the same symmetry (topology), so their total dipole moment 
is close to zero. The ZnO part of the  slab is two lattice constants $c_{ZnO}$ thick on both sides and 12~{\AA}  of vacuum layer have been added to separate the 
slabs in {\textit z} direction. Additionally, we have studied the bulk spinel
properties, using the $12\times12\times12$ {\textit k}-point grid, and its clean (111) surface within the same method. 
We simulated the Co-terminated and O-terminated spinel (111) surfaces using the slabs, created for the interface model, but with ZnO layers removed and followed by subsequent 
relaxation over all coordinates. 

\section{Results and discussion}
We constructed the interface and surface models assuming that the secondary phase preserves bulk spinel crystal structure with the corresponding bulk constant. Our assumption is based on the comparison with the Raman 
spectra calculated \cite{Mock2016} and measured \cite{29, Phan2012} for both \spinel and $\rm Zn_{1-x}Co_xO$. In general, the symmetry of the bulk 
spinel unit cell is described by point group $\Gamma(O_h^7)$,\cite{Rousseau1981} and therefore the phonon normal modes near the Brillouin zone center 
may be obtained by the decomposition $\Gamma(O_h^7) = A_{1g}+E_g+3F_{2g}+5F_{1u}+2A_{2u}+2E_u+2F_{2u}$. Here $A_{1g}$, $E_{g}$ and triple degenerated $3F_{2g}$ modes 
are Raman active. We calculated the frequencies of these phonon modes for bulk spinel with lattice constant $a_{spinel}=8.084$~{\AA} (corresponding 
to the case of normal spinel secondary phase and compressed ZnO at the interface), and ``stretched" spinel using experimental ZnO bulk constant, which leads to $a_{spinel}=9.191$~{\AA}, using density-functional perturbation theory.\cite{Baroni2001} The PBE pseudopotentials were selected in norm-conserving form, the wave function 
expansion cutoff of 80~Ry and $4\times4\times4$ k-point grid for Brillouin zone integration were adopted for this calculations. The calculated and measured Raman frequencies are 
collected in Table~\ref{tab:raman}. It demonstrates that the calculated Raman spectra are very sensitive to the choice of the lattice constant. Frequencies obtained in both LDA and GGA approximations 
for normal spinel are comparable with measured ones, while those calculated for ``stretched" spinel are found to be significantly lower and are not observed experimentally. Moreover,
XRD measurements of $\rm Zn_{1-x}Co_xO $ \cite{Phan2012} do not indicate a presence of any other structures beside of ZnO and $\rm ZnCo_2O_4$. The above comparison of the theoretical and experimental frequencies is in favor of using the bulk \spinel constant when modeling the interface with ZnO. 

The calculated lattice constant for bulk spinel $a_{spinel}=8.147$~{\AA}, and the corresponding interplanar A-B spacing $d_{111}=2.351$~{\AA} are 
overestimated by only 0.8\% compared to the experimental values of $a_{spinel}=8.084$~{\AA} and $d_{111}=2.333$~{\AA}, respectively. Therefore, we used the experimental spinel bulk constant. 

\begin{table}[h]
\centering
\caption{\label{tab:raman} Raman-active bulk phonon modes of \spineln, $\rm cm^{-1}$.Two last lines show the frequencies, calculated in this study }
\begin{ruledtabular}
\begin{tabular}{llllll}
&$F_{2g}$&$E_{g}$&$F_{2g}$&$F_{2g}$&$A_{1g}$\\
\hline%
\spinel [Ref.\cite{29}]&194.4&482.4&521.6&618.4&691.0 \\     
$\rm Zn_{1-x}Co_xO$ [Ref.\cite{Phan2012}]&&486&524&623&710 \\     
\spinel LDA [Ref.\cite{Mock2016}]&192&480&511&589&644 \\     
\spinel GGA &187.0&463.9&502.8&574.5&631.7 \\
stretched \spinel GGA&62.3&175.4&236.8&325.2&383.8\\
\end{tabular}
\end{ruledtabular}
\end{table}

As mentioned in Sec.3, the unit cell of the spinel (111) plane in the slab construction is hexagonal, and therefore 4 unit cells of ZnO (also hexagonal) 
are needed to match one spinel unit cell. Consequently, the planar lattice constant of adjacent ZnO $a_{ZnO}=2.88$~{\AA} is scaled to the spinel lattice 
constant and cannot be optimized separately. However, the interplanar distances (in {\textit z} direction) are optimized, for both the spinel and the wurtzite regions of the interface. 
Therefore, the calculated value of the interplanar spacing at the spinel region of the interface becomes $d_{111}=2.387${~\AA}, which is about 2\% larger than the experimental bulk interplanar 
distance, while the lattice constant, calculated for ZnO regions, $c_{ZnO}=5.52$~{\AA}, which is about 5\% above the corresponding experimental bulk 
value of 5.27~{\AA}. These relaxations absorb part of the stress due to the lattice mismatch between spinel and wurtzite. 
The optimized supercells of \inter interfaces are shown in Fig.~\ref{fig:struct}. Since there are no dangling bonds at the interfaces and all the ions are located in such 
a way that the bulk crystalline symmetry is preserved, no significant modifications in a topology of adjacent atomic layers were found during the 
relaxation. In the case of surfaces, there exist four possibilities: B- or A-termination with Co or O top layer. The B-terminated sample with Co top layer 
demonstrates  atomic reordering: the oxygen atom of the second layer  (O atom circled by dashed line in Fig.~\ref{fig:struct}a) moves in {\textit z}-direction to be 
in the same plane as the Co atoms of first layer. Such reordering occurs in \cobn-terminated surface only: A-terminated surfaces with both Co and O top layers 
and B-terminated with O top layer demonstrate stable surface topology with no significant changes of the overall atomic positions compared  to those in the interfaces. 
We have also performed geometry optimization for 
9 atomic layer - thick \spinel slabs, and found that the results are practically identical to the case, considered in Fig.~\ref{fig:struct}.	

\begin{table}[h]
\caption{\label{tab:moments} Magnetic moments $\mu$ (in the units of $\mu_B$) and charges $\rho$ of Co ions (in a.u.), calculated using L{\"o}wdin charge analysis, for octahedral surfaces and interfaces.}
\begin{ruledtabular}
\begin{tabular}{cccccc}
&&Co-terminated&O-terminated&ZnO&Bulk \\
&&surface&surface&interface&\\ 
\hline%
\cob&$\mu$&2.33&0.71&0.21&0.0 \\     
&$\rho$&0.94&1.17&1.07&1.02 \\ 
\coa&$\mu$&2.45&2.48&2.46&2.59\\
&$\rho$&1.16&1.23&1.21&1.22\\
\end{tabular}
\end{ruledtabular}
\end{table}

It has been discussed above that in the bulk spinel \cob ions are non-magnetic due to the large splitting between $t_{2g}$ and $e_g$ orbitals, caused by the 
presence of octahedral crystal field. Since this symmetry is broken at the surface or interface, the electrons could occupy $t_{2g}$ and $e_g$ orbitals 
in different order, leading to the changes in magnetic properties, as reported in [\onlinecite{34},\onlinecite{43}]. It is important to stress that similar symmetry changes are typical for other \spinel 
interface orientations, therefore the results  for the \spineln(111) surface, considered here, should reflect general trends in the interface induced magnetism origin.  To quantify these changes, we calculated and 
compared the magnetic moment of Co ions for different interface and surface systems using a L{\"o}wdin charge analysis. Table ~\ref{tab:moments} shows the largest values 
of magnetic moments, calculated for the bulk \spineln, interfaces and surfaces, both Co- and O-terminated. The magnetic moments for \cob ions are 
calculated for the top layer and for \coa ions in the second layer of octahedral interface or surface. The deviation of the magnetic moment of the same 
ion type on different sites is relatively small $\sim0.02~\mu_B$ for all systems, so such values are reflecting the general physical picture.

\begin{figure}
\includegraphics{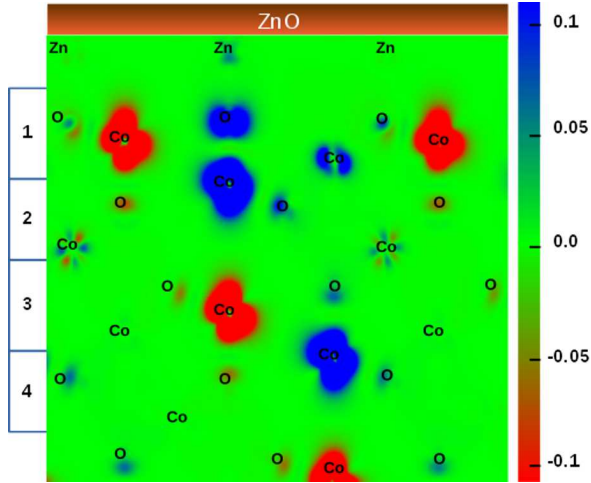} 
\caption{Spin density distribution for octahedral interface, plotted over vertical $\rm (1\bar 10)$ plane. The scale, shown at the right, has units of $\mu_B$. At the left, 1, 2, 3 and 4 stand for the \spinel interface layer numbers.  Chemical symbols indicate the positions of corresponding ions.}
\label{fig:tet_den}
\end{figure}

The calculated magnetic moment of \coa ions in bulk spinel is $2.59~\mu_B$ is  slightly smaller in the case of all considered surfaces, as it is seen from 
Table~\ref{tab:moments}. Instead, while the magnetic moment of \cob ions is zero in the bulk, it is non-vanishing in the case of the surface. The 
largest magnetic moment of $2.33~\mu_B$ occurs at the Co-terminated surface, where the bulk symmetry is broken and ion coordination number is reduced 
at the most. If the surface is O-terminated, the magnetic moment of \cob ions reduces to $0.71~\mu_B$, while the external oxygen atoms receive a 
magnetic moment of $0.34~\mu_B$ due to a strong polarization of the {\textit p}-orbitals. The charge, calculated for \cob ions in the bulk, is about 0.2~a.u. larger than that one of \coan, 
as shown in Table~\ref{tab:moments}. These values differ slightly for all of the systems under study,  and, in general, we have to introduce new oxidation state types for Co ions in 
interfaces and surfaces. However, in our calculations the charge of \cob is always larger then that of \coan, and  
this fact allow us, for the sake of simplicity, to use explicit ``bulk'' notations \coa and \cob for corresponding ions in all of systems.
The spin density distribution for tetrahedral interface is shown in Fig.~\ref{fig:tet_den}. The blue and red colors regions around the \coa ions of the Layers 1 and 3 
indicate the presence of magnetic moment, comparable to that in the bulk. \cob ions of the Layers 3 and 4  are completely bare, that is spin compensated,
but receive small magnetic moment in Layer 2, which
becomes noticeably larger at the interfacing layer. Similarly to the case of O-terminated surface, one of oxygen ions acquires a magnetic moment of $0.22~\mu_B$, 
as indicated by blue colour. Obviously, such a magnetic ordering corresponds to AFM state: we calculated the total energy for the different spin orientations, and 
for this tetrahedral interface the difference between the energies of ferromagnetic (FM) and antiferromagnetic (AFM) states is $\rm E_{FM}-E_{AFM}= 94~meV$. For the octahedral interface FM state is energetically preferable
and the difference in energy between FM and AFM states is -23~meV. 
In general, the lowest total energy is found for octahedral interface with FM magnetic ordering. 

More accurate method to estimate the relation between the interface type and magnetic ordering is to calculate the formation energy. Such an approach, however, requires the knowledge of the chemical potentials of participating ions. To the best of our knowledge such problem has not been solved yet: the main challenge is to properly find these potentials for ions in different oxidation states.

It is worth to note, that the magnetic moments were calculated for the relaxed systems while keeping the $\rm C_{3v} $ symmetry intact. 
If this symmetry is broken (for instance, for differently oriented interfaces or when the initial deviations from equilibrium positions are different for symmetry equivalent atoms, or due to defects), 
the corresponding magnetic moments might differ slightly. Nevertheless, the general picture should remain the same: \cob ions are gaining the non-zero magnetic 
moments both at the surface and interface, in contrast to the bulk case. Therefore, the magnetic effects, discussed above, should also be present for differently oriented parts of the \spinel inclusions.

\begin{figure*}
\includegraphics{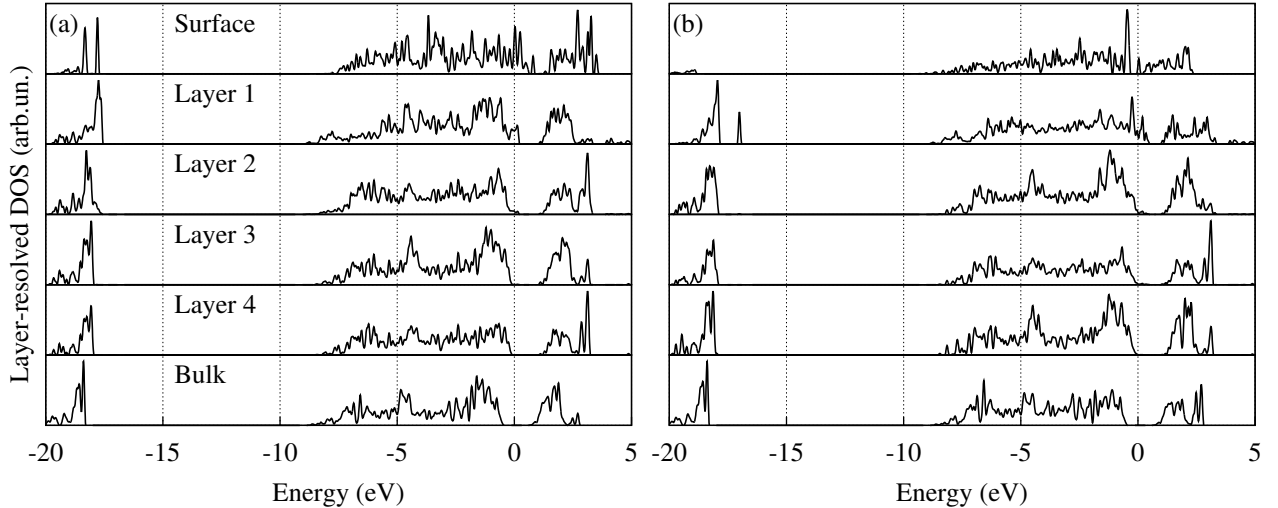} 
\caption{Layer-resolved DOS of (a) octahedral and (b) tetrahedral spinel (111) surfaces and (111) spinel / (0001) wurtzite interfaces. From top to bottom: top layer of 
spinel (111) surface with no ZnO cap; layers 1 to 4 of the spinel structure close to the spinel/wurtzite interface and the layer-resolved DOS for 
the (111) plane of bulk spinel (see discussions in the text). } 
\label{fig:lrpdos}
\end{figure*}

As it is known, a presence of the dangling bonds leads to additional surface states, observable in the density of states (DOS). Formation of the 
interface between two different materials is  also responsible for the interface states, localized close to the boundary between the two materials. The surface or interface formation 
causes the charge redistribution and change in the corresponding magnetic properties. To demonstrate this we first calculated the spin-averaged layer-resolved DOS (LRDOS) for all 
systems under investigation, as shown in Fig.~\ref{fig:lrpdos}. For the bulk spinel, the planes that pass through the Co-ions of corresponding charge 
state (A or B type) were used as for LRDOS calculations. All LRDOSs there are aligned in such a way that the highest filled states (Fermi level) are at 
zero energy. For the top layer of Co-terminated surfaces, there is clear evidence of such surface states present in the DOS (upper panels on Fig.~\ref{fig:lrpdos},
denoted ``surface''). 
It contains a lot of features, not present in the bulk, and such a picture, in principle, is typical for all the considered surfaces with the dangling bonds. There is a notable difference 
in Co-terminated surface DOS for octahedral and tetrahedral termination at the region -18~eV, due to the oxygen atom shift from layer 2 and now 
belonging to the top layer of octahedral system. On the other hand, in tetrahedral system top layer consists of Co atoms only. LRDOS of O-terminated 
surfaces (not shown) demonstrates no noticeable difference, compared to the Co-terminated surface for both tetrahedral and octahedral coordinations. 
In this case, for both coordinations, the bonds of surface Co atoms now are passivated by oxygen atoms and are not broken anymore. This means, that 
there are also other factors responsible for the formation of the inside band-gap states. Such a situation is also observed in the case of interface. 
One can see from Fig.~\ref{fig:lrpdos} (panels denoted ``Layer n'' with n=1, 2, 3, and 4), that LRDOS for the first layer demonstrates surface-like 
states inside the band-gap, close to the top of the valence band. For the internal layers these surface-like states are decaying with depth, and almost 
disappearing at Layer 4. Corresponding LRDOS becomes bulk-like, both for octahedral and tetrahedral coordinations, as seen from the comparison 
between LRDOS of Layer 4 and those denoted ``Bulk'' on Fig.~\ref{fig:lrpdos}. 

\begin{figure}
\includegraphics{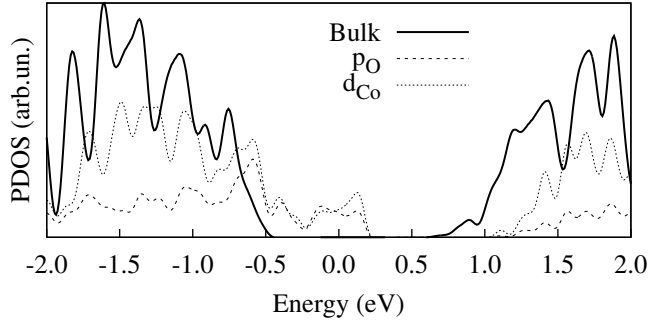} 
\caption{Projected LRDOS for top layer
 of octahedral interface and octahedral plane of bulk spinel. The plot demonstrates domination of the {\textit p}- and {\textit d}-components of the wave function, 
while the {\textit s}-state contributions can be neglected.} 
\label{fig:spd}
\end{figure}

Comparing LRDOS calculated for surface and interface, one can conclude, that although each Co-ion at the interface layer keeps the symmetry of the 
bulk crystalline environment, the physical properties of the interface region is closer to the surface, rather than to the bulk. To understand the origin of 
the surface-like states in the band-gap, we calculated the LRDOS of the octahedral interface, projected onto atomic wavefunctions of corresponding Co 
atom ({\textit s} and {\textit d} orbitals) and O atom ({\textit s} and {\textit p} orbitals), localized at the octahedral interface, as shown in  
Fig.~\ref{fig:spd}. For convenience, we plotted there also LRDOS for A-plane of bulk spinel. As it can be seen, surface-like states originate 
predominantly from O 2{\textit p} states and Co 3{\textit d} states, while the contribution of {\textit s}-states of both Co and O is negligibly small 
here. Similar conclusions for the origin of the surface states in the tetrahedral systems have been also obtained. From this we conclude that charge state 
of Co-ion is not decisive in defining the surface or interface magnetism since in both of cases {\textit p}-orbitals of O-atoms make the same contribution 
into DOS. Moreover, from the band structure calculation we see that the partial occupied states are common for all of the surfaces and interfaces under 
investigation. This demonstrates the metal-like electronic structure, in contrast to the bulk spinel, which appears semiconducting in the simulations 
even when larger smearing parameters in the Brillouin zone integration are used.

\section{Conclusion}   
  We investigated the origin of the surface/interface magnetism  of the cobalt oxide \spinel surfaces and their  interfaces with zinc 
oxide ZnO. In particular, we studied the structural, electronic and magnetic properties using the model systems such as ZnO(0001)/\spineln(111) interfaces, 
\spineln(111) surfaces for A-type and B-type terminations and bulk spinel. 
It is shown that while the magnetic moment of \cob ions is zero in the bulk, it does not vanish at the interface or surface, where its value becomes comparable with the magnetic 
moment of \coa due to the created imbalance in the electron distribution. The calculated LRDOS demonstrates that 
although Co ions at the interface have the same neighboring atoms as in bulk spinel, their DOS exhibit the surface-like nature, arising from polarized 
Co 3{\textit d} and O 2{\textit p} orbitals of the interfacing layer. In all cases, interface or surface, A- or B-type termination, we observe 
metallic-like states,  localized at the surface or interface, and which are responsible for the surface/interface magnetism. Whereas the magnetic 
order is antiferromagnetic in the bulk spinel at low temperature, the metallic surface/interface states indicate the possibility of a 
ferromagnetic order at the surfaces or interfaces. The proposed mechanisms offer possible interpretation of  the experimental observation of the net magnetic moment in certain Co 
doped ZnO with high Co concentrations. 

\begin{acknowledgments}
We thank O. Kolomys and V. Strelchuk for usefull discussions. The work was supported by Science for Peace and Security Program (the grant NATO NUKR.SFPP 984735). 
I. Kupchak acknowledges EC for the RISE Project CoExAN GA644076 within HORIZON2020 program. The CPU time was provided by the Shared Hierarchical Academic 
Research Computing Network (Sharcnet) of Ontario, Canada.
\end{acknowledgments}


%

\end{document}